\documentstyle[floats,tighten,epsf,prl,aps]{revtex}

\def\be{\begin{equation}}
\def\ee{\end{equation}}

\def\bea{\begin{eqnarray}}
\def\eea{\end{eqnarray}}
\def\bml{\begin{mathletters}}
\def\blea{\begin{mathletters}\begin{eqnarray}}
\def\elea{\end{eqnarray}\end{mathletters}}

\begin{document}
\draft
\wideabs{
\title{Radiation from cosmic string standing waves}

\author{Ken D.\ Olum\footnote{Email address: {\tt kdo@alum.mit.edu}}
and J.\ J.\ Blanco-Pillado\footnote{Email address: {\tt
jose@cosmos2.phy.tufts.edu}}}

\address{Institute of Cosmology \\
Department of Physics and Astronomy \\
Tufts University \\
Medford, MA 02155}

\date{October 1999}

\maketitle

\begin{abstract}%
We have simulated large-amplitude standing waves on an Abelian-Higgs
cosmic string in classical lattice field theory.  The radiation rate
falls exponentially with wavelength, as one would expect from the
field profile around a gauge string.  Our results agree with those of
Shellard and Moore, but not those of Vincent, Antunes, and Hindmarsh.
The radiation rate falls too rapidly to sustain a scaling solution via
direct radiation of particles from string length.  There is thus
reason to doubt claims of strong constraints on cosmic string theories
from cosmic ray observations.
\end{abstract}


}
\def\thefootnote{\fnsymbol{footnote}}
\footnotetext[1]{Email address: {\tt kdo@alum.mit.edu}}
\footnotetext[2]{Email address: {\tt jose@cosmos2.phy.tufts.edu}}
\def\thefootnote{\arabic{footnote}}
\narrowtext
\paragraph*{Introduction}
Cosmic strings are one-dimensional topological defects which may have
been created by a phase transition in the early universe
\cite{Kibble76}. (For reviews see \cite{Alexbook,Kibble95}.)
As the universe evolves, intercommutations between long strings
produce oscillating loops.  In the standard scenario, these loops lose
energy by gravitational radiation and eventually disappear.  This
produces a scaling solution where the average distance between strings
is a constant fraction of the Hubble length.  Most of the energy in the
string network is emitted as gravitational waves, which we cannot
observe, and only a small fraction appears as high-energy particles.

However, in a recent paper \cite{Hind98}, Vincent, Antunes, and
Hindmarsh claim that energy in a string network is lost by direct
particle emission from long strings, rather than in gravitational
waves.  To back up this claim they study large-amplitude sinusoidal
standing waves, and claim that the energy emission rate is sufficient
to explain scaling behavior with the great majority of the energy
emitted as particles.  Moore and Shellard \cite{Shellard98} found that
the emission rate fell exponentially with wavelength, but their
amplitudes \cite{Shellardlumps} were much less than those of
\cite{Hind98}.  Furthermore, the range of wavelengths in which
\cite{Shellard98} saw exponential fall off had no overlap with the
wavelengths studied in \cite{Hind98}.

Here we simulate the same large-amplitude waves as in \cite{Hind98} 
and cover part of the same wavelength range, but we come to a different
conclusion.  In our simulations, the energy emission rate declines
exponentially with wavelength, and thus cannot account for the large
direct-emission rate claimed in \cite{Hind98}.

\paragraph*{Model}
We work with the Abelian-Higgs model, which produces local strings
with no massless degrees of freedom in the vacuum.  The Lagrangian is
\be\label{eqn:Lagrangian}
{\cal L} = D_\mu\bar \phi D^\mu\phi -{1\over 4} F_{\mu\nu} F^{\mu\nu}
-{\lambda\over 4}(|\phi|^2-\eta^2)^2\,.
\ee
We work with units such that $\eta = 1$ and $e = 1$, and we use the
``critical coupling'' regime in which $\beta =\lambda/(2e^2) = 1$ so
that in our units $\lambda = 2$.  As in \cite{Hind98}, we study
strings whose initial core position is given by a sinusoidal wave $y =
A \cos kx$, with amplitude $A =\lambda/2 =\pi/k$ \cite{sinusoids}.

\begin{figure}
\begin{center}
\leavevmode\epsfbox{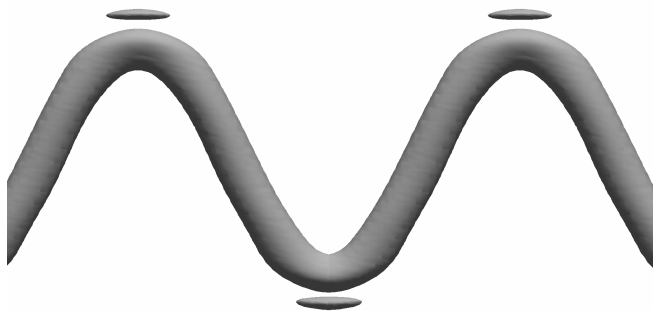}
\end{center}
\caption{A surface of constant energy density shows a sinusoidal wave
on a cosmic string and the energy emitted from the high curvature
regions.}
\label{fig:energy-from-tips}
\end{figure}

\begin{figure}
\begin{center}
\leavevmode\epsfbox{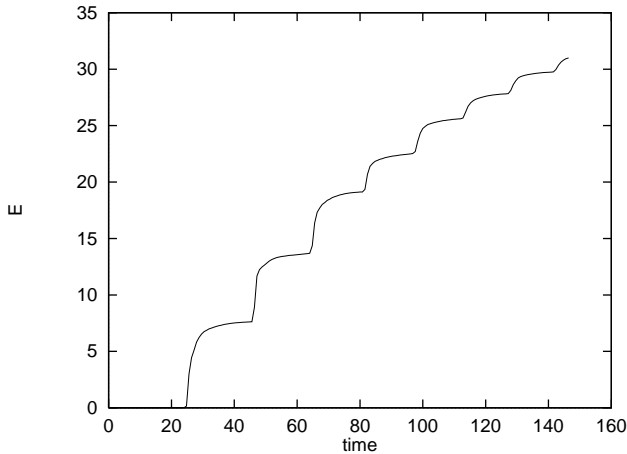}
\end{center}
\caption{Energy emitted from one wavelength of a standing wave.  The
radiation appears as a series of bursts.}
\label{fig:oscillation-energy}
\end{figure}

A preliminary investigation shows that emission from a standing wave
is not uniform in time, but rather consists of a series of bursts
emitted when the string is momentarily stationary with large amplitude
waves in its position.  Figure \ref{fig:energy-from-tips}
shows a snapshot of the energy density around a string at one such point,
and Fig.\ \ref{fig:oscillation-energy}
shows a plot of the energy emitted over several oscillations.

It thus appears that standing wave radiation is
akin to radiation from cusps, and results from the overlap of the
tails of the the string fields.  We use such a model below to compute
a theoretical expectation of the dependence of radiation rate on
wavelength.

\paragraph*{Expectations}

Vincent, Antunes, and Hindmarsh \cite{Hind98} argued as follows: in a
scaling network, the distance between strings, $\xi$, scales with the
Hubble distance, which is proportional to time.  In a volume $\xi^3$
there will be string length roughly $\xi$, so the energy density in
the string network is $\rho =\mu/\xi^2$, and thus $\dot\rho =
-2\mu\dot\xi/\xi^3$.  Since $\dot\xi$ is a constant,
$\dot\rho\xi^3$ is constant.  As a model they used a sinusoidal
standing wave with wavelength $\lambda$ in a box of volume $\lambda^3$.
They expected $\dot\rho\lambda^3$ to be constant.  If we let $E$ be
the energy of a single wavelength of the string, then $E
=\rho\lambda^3$, and thus $\dot E$ is independent of
$\lambda$.  If we let $P_L$ be the power per unit length radiated from
the standing wave, we need
\be\label{eqn:vah-power}
P_L\propto\lambda^{-1}
\ee
to sustain a scaling network from energy emission of this type.

In contrast, analyzing the fields around the string would lead to a
different conclusion.  A straight, static string is topologically
stabilized in a minimum-energy configuration, and so cannot radiate.
If the string is curved, then there is the possibility for radiation,
but since the fields fall off exponentially toward the vacuum at large
distances from the string, one would expect the amount of radiation to
be suppressed by an exponential factor depending on the radius of
curvature, $R$.  This seems in keeping with Fig.\
\ref{fig:energy-from-tips}, which shows the radiation coming from the
points of maximum curvature.

As a specific model, one can imagine that an element of momentarily
stationary curved string gives up an amount of energy proportional to
$\exp (-\alpha R) dl$, where $\alpha$ is a constant of $O (1)$ and
$dl$ is the length of the element of string.  The total energy
emission is then
\be\label{eqn:energy-integral}
E\propto\int e^{-\alpha R} dl\,.
\ee

For a sinusoidal wave, $y = A\cos kx$, the radius of curvature is
\be
R ={(1+A^2k^2\sin^2kx)^{3/2}\over Ak^2\cos kx}\,.
\ee
We will consider the region around one of the peaks of the sinusoid.
The energy emission is dominated by the region where $x$ is near zero,
so that $R$ is small.  In this regime, we can approximate
\be
R\approx{(1+A^2k^4x^2)^{3/2}\over Ak^2 (1-k^2x^2/2)}
\approx{1\over Ak^2} +\left({3\over 2} Ak^2+{1\over 2A}\right) x^2\,.
\ee
In our case, we are going to consider a fixed ratio of amplitude to
wavelength, $A =\lambda/2 =\pi/ k$ so we get
\be
R\approx{\lambda\over 2\pi^2} +{3\pi^2+1\over\lambda} x^2\,.
\ee
We can now do the integral of Eq.\ (\ref{eqn:energy-integral}),
approximating $dl =\sqrt{1+\pi\sin^2kx}\, dx$ by just $dx$, and
extending the limits of integration to infinity, to get
\be
E\propto\sqrt{\lambda}\, e^{-\alpha\lambda/(2\pi^2)}\,.
\ee

Since we keep the amplitude a fixed multiple of the wavelength, the
period of the standing wave is just proportional to $\lambda$.  If we
consider a half wavelength of string, it emits bursts of energy $E$
twice per cycle, so the power is $\propto\lambda^{-1/2}
e^{-\beta\lambda}$, and the power per unit length is
\be\label{eqn:theoretical-power}
P_L\propto\lambda^{-3/2} e^{-\beta\lambda}\,.
\ee

\paragraph*{Simulation}

The simulation is based on a lattice action, as described in
\cite{JJKDO98.1}.  However, in the present case we have used a
different lattice spacings in the 3 cardinal directions.  The maximum
speed of the string is quite large, and leads to a Lorentz contraction
of the field profile in the direction of motion.  To accurately
represent the fields, the lattice spacing should be proportional to
$1/\gamma_i = \sqrt{1-v_i^2}$, where $v_i$ is the component of the
string velocity along axis $i$.  In the $z$ direction, where there is
no motion, we have used a lattice spacing of 0.33, which seems to be
the largest that gives reliable results.  The corresponding spacings
in the $x$ and $y$ directions are 0.31 and 0.10 respectively. The 
Courant condition requires
 $\Delta t < {({\Delta x}^{-2} +{\Delta y}^{-2} + 
{\Delta z}^{-2})}^{-1/2} \approx 0.09$, and in our simulations we 
use $\Delta t = 0.08$.

To extract the energy which is emitted by the string, we have used
absorbing boundary conditions on the $y$ and $z$ faces and accumulated
at each step the amount of energy that they absorb.  The conditions are
\bml\label{eqn:boundary-conditions}\bea
n_i D_i \phi &=& - D_t \phi\,,\\
{\bf E}_T &=& - {\bf n} \times {\bf B}\,,
\elea
where $D$ denotes the covariant derivative, ${\bf n}$ the outward
 normal unit vector on each boundary, and 
${\bf E}_T\equiv{\bf E} - {\bf n} ({\bf E} \cdot {\bf n})$ the
transverse component of the electric field. This corresponds to 
the zeroth-order absorbing boundary condition for free
 electromagnetism\cite{Absorbing97}.
 The energy flowing into the boundary is
\be \label{eqn:e-m-conserved}
E_{absorbed} =  \int_{\Omega}  {\bf S} \cdot d{\bf n}
\ee
where $\Omega$ is the boundary surface, and ${\bf  S}$ is the Poynting
vector, given by
\be
S_j = - D_t \bar \phi D_j \phi - D_t \phi D_j \bar \phi + 
({\bf E} \times {\bf B})_j\,.
\ee
Using Eqs.\ (\ref{eqn:boundary-conditions}), we can rewrite 
Eq.\ (\ref{eqn:e-m-conserved}) as
\be
E_{absorbed} = \int_{\Omega} (2 |D_t \phi|^2 + |{\bf E}_T|^2 ) d \Omega\,,
\ee
which is easily computed.

To produce the sinusoidal waves, we use the same technique that we
used to generate cusps in \cite{JJKDO98.1}, i.e., we create two
traveling wiggles on a straight string that will combine to produce
the desired sinusoidal form.  (In the Nambu-Goto approximation, the
resulting sinusoid would be exact; in our case there will be a
distortion of the shape due to the dynamics that occur before it is
formed, but this effect will be small because the wiggles out of which
the sinusoid forms are not themselves strongly curved.)  The initial
field configuration for a moving wiggle is known exactly, from a result
of Vachaspati \cite{Tanmay90}.  The advantage of this technique is
that it does not require the use of relaxation, as is necessary in
other field theory simulation schemes \cite{Hind98,Shellard98}.

The two original wiggles will overlap to form a single wavelength of
the standing wave, from one minimum of $y$ to the next.  At this point
it is possible to change to periodic boundary conditions in the $x$
direction, so that the straight part of the string is removed, and we
are left with a single wavelength of standing wave in a periodic box.
This technique was used to produce Figs.\ \ref{fig:energy-from-tips}
and \ref{fig:oscillation-energy}, but it is not an accurate method for
extracting the radiation rate, because energy coming from different
burst is not clearly separated by the time it reaches the boundaries.
When each burst of radiation is emitted, the string changes its shape
and amplitude and its subsequent evolution does not correspond to a
constant amplitude sinusoid any more.  Of course, for large enough
$\lambda$ this would not matter, but for the wavelengths in the range
of our simulation, it makes a significant difference.

To avoid this problem, we allow the original wiggles to pass by each
other beyond the point of the overlap, so that they generate just a
single burst, and then separate.  The place from which the burst is
emitted is at the center of the overlapping region, and the string
near that point has been following the same evolution as in a real standing
wave for a half period, so we feel that this burst accurately
represents a single burst of a standing wave oscillation.

\begin{figure}
\begin{center}
\leavevmode\epsfbox{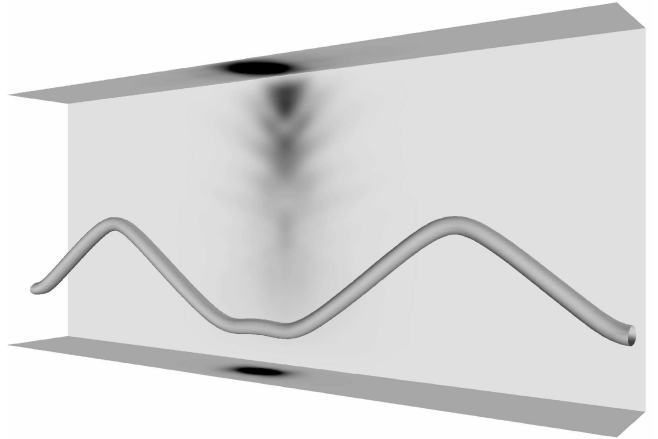}
\end{center}
\caption{Absorption of a burst of energy.  The two wiggles have passed
through each other and are now separating.  Three sides of the
bounding box are shown, with darker shades showing areas of greater
energy absorption.}
\label{fig:absorption}
\end{figure}

Using the expressions given above for the Poynting vector on the
boundaries, we can compute the energy absorbed at each time step in
our simulation.  The energy absorption on the box faces is shown in
Fig.\ \ref{fig:absorption}.
Integrating all the energy absorbed
from the first burst of radiation we can compute the power
emitted by a sinusoidal standing wave.

\begin{figure}
\begin{center}
\leavevmode\epsfbox{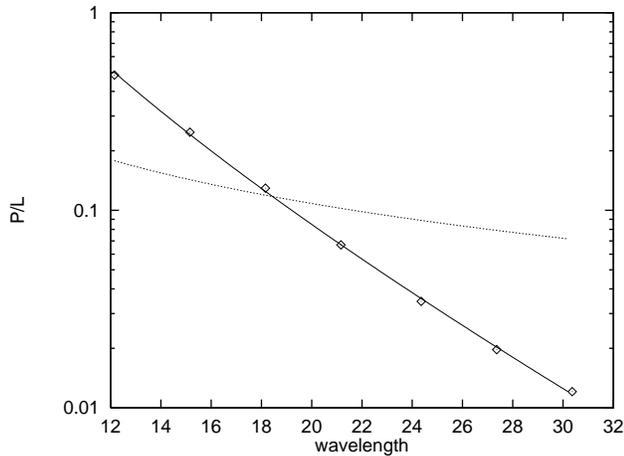}
\end{center}
\caption{Power per unit length from simulation (points) compared with
exponentially decaying model (solid line) and linearly decaying model
of \protect\cite{Hind98} (dotted line).}
\label{fig:results}
\end{figure}

We have repeated this procedure for different values of the wavelength
ranging from $\lambda = 12$ to 30, in natural units.  In Fig.\
\ref{fig:results}
we plot the power emitted per unit length and compare it with
theoretical predictions from Eqs.\ (\ref{eqn:theoretical-power}) and
(\ref{eqn:vah-power}).  We see that the exponentially decaying model
fits quite well.  The line shown has the form
\be
P_L\propto\lambda^{-3/2} e^{-2.56 R_0}
\ee
with $R_0\equiv\lambda/(2\pi^2)$, but we do not have sufficient
accuracy to confirm the exponent of $\lambda$ or the exact constant.
A curve with $\lambda^{-2}$ or $\lambda^{-1}$ and somewhat different
constant would fit equally well.  On the other hand, the form of
\cite{Hind98} does not fit at all.

\paragraph*{Discussion}

We have simulated large-amplitude\break standing waves on local cosmic
strings, and found an exponential decrease in radiation power with
increasing wavelength.  Our technique proceeds from separated wiggles
with exact initial conditions, and we have used quite a small lattice
spacing as compared to other authors, so we feel that our results are
reliable.

The constant in the exponential gets its dimensions from the string
thickness ($10^{-30}\text{cm}$ for a GUT scale string), and thus for
waves of any reasonable cosmological size, the radiation is utterly
negligible.  One could in principle imagine that strings in
cosmological networks still have excitations at wavelengths comparable
to their thickness, but this does not seem reasonable.  Such wiggles
will be rapidly smoothed out by gravitational radiation, and there is
no mechanism for regenerating them at such small scales.  Thus we
conclude that direct radiation of particles from string length cannot
play a significant role in the production of cosmic rays or the
maintenance of a scaling network.  As a result, cosmic ray
observations do not rule out field theories that admit cosmic
strings, as claimed in \cite{Hind98,Wichos98,Bhattacharjee:1997in}.

\paragraph*{Acknowledgments}

We would like to thank Alex Vilenkin and Xavier Siemens for helpful
conversations, and Southampton College, particularly Arvind Borde and
Steve Liebling, for the use of their computer facilities.  This work
was supported in part by funding provided by the National Science
Foundation.  J. J. B. P. is supported in part by the Fundaci\'on Pedro
Barrie de la Maza.


\begin{thebibliography}{10}
\bibitem{sinusoids}
Note that unlike the situation with a linear system such as a one-dimensional
  scalar field, there is nothing special about a sinusoid. In particular, the
  oscillation period of a sinusoidal standing wave on a string depends on the
  length along the string, rather than $\lambda$, which is the spatial distance
  occupied by one wavelength.
\end{thebibliography}
\end{document}